\pdfoutput=1               % force arXiv to use pdfLaTeX :contentReference[oaicite:5]{index=5}
\documentclass[numbered,nowordcount]{trbunofficial} % turn off internal counter
\usepackage{graphicx}

\AuthorHeaders{Chen, Chen, Li, Yang and Feng}
\title{Impact Analysis of Inference Time Attack of Perception Sensors on Autonomous Vehicles}

\author{%
  \textbf{Hanlin Chen*, Corresponding Author, ORCID: 0000-0001-6508-7715}\\
  Buildings and Transportation Science Division\\
  Oak Ridge National Laboratory, Oak Ridge, USA,37932\\
Email: chenh1@ornl.gov \\
  \hfill\break% this is a way to add line numbering on empty line
  \textbf{Simin Chen}\\
  Department of Computer Science,  The University of Texas at Dallas
\\
  800 W Campbell Rd, Richardson, TX 75080 \\
  Email: sxc180080@utdallas.edu\\
  \hfill\break%
\textbf{Wenyu Li}\\
Department of Civil and Environmental Engineering, The Hong Kong University of Science and Technology\\
2006 Xiyuan Ave, HongKong, China 611731 \\
Email: wenyu.li.cn@gmail.com\\
  \hfill\break%
    \textbf{Wei Yang, Ph.D.}\\
    Department of Computer Science, The University of Texas at Dallas \\
    ECSS 4.225, 800 W. Campbell Rd., Richardson, TX 75080\\
    Email: wei.yang@utdallas.edu\\
  \hfill\break%
  \textbf{Yiheng Feng, ORCID: 0000-0001-5656-3222 }\\
  Lyles School of Civil Engineering, Purdue University \\
550 Stadium Mall Drive, West Lafayette, IN 47907 \\
Email: feng333@purdue.edu 
}
\usepackage{booktabs}
\usepackage{amsfonts}
\usepackage{graphicx}
\usepackage{hyperref}
\usepackage{textcomp}
\usepackage{listings}
\usepackage{adjustbox}
\usepackage{xspace}    % sticks a sane space after a command
\usepackage{multirow}
\usepackage{multicol}
\usepackage{xcolor} 
\usepackage{amsmath}
\usepackage{tcolorbox}
\usepackage{subfig}
\usepackage[linesnumbered,boxed,ruled]{algorithm2e}
\usepackage{listings}

\usepackage{wasysym}
\usepackage{algorithmic}

\definecolor{codegreen}{rgb}{0,0.6,0}
\definecolor{codegray}{rgb}{0.5,0.5,0.5}
\definecolor{codepurple}{rgb}{0.58,0,0.82}
\definecolor{backcolour}{rgb}{0.95,0.95,0.92}

\lstdefinestyle{mystyle}{
  backgroundcolor=\color{backcolour},   commentstyle=\color{codegreen},
  keywordstyle=\color{magenta},
  numberstyle=\tiny\color{codegray},
  stringstyle=\color{codepurple},
  basicstyle=\ttfamily\footnotesize,
  breakatwhitespace=false,         
  breaklines=true,                 
  captionpos=b,                    
  keepspaces=true,                 
  numbers=left,                    
  numbersep=5pt,                  
  showspaces=false,                
  showstringspaces=false,
  showtabs=false,                  
  tabsize=2
}

\begin{document}
\maketitle
%\fi
\section{Abstract}

As a safety-critical cyber-physical system, cybersecurity and related safety issues for Autonomous Vehicles (AVs) have been important research topics for a while. Among all the modules on AVs, perception is one of the most accessible attack surfaces, as drivers and AVs have no control over the outside environment. Most current work targeting perception security for AVs focuses on perception correctness. In this work, we propose an impact analysis based on inference time attacks for autonomous vehicles. We demonstrate in a simulation system that such inference time attacks can also threaten the safety of both the ego vehicle and other traffic participants.

\hfill\break%
\noindent\textit{Keywords}: Autonomous Vehicle, Cybersecurity, Perception security, Cyber-physical sytstem
\newpage
%%%%%%%%%%%%%%%%%%%Main part of the paper
\section{Introduction}

% \TODO{From VehicleSec ver, ADJUST!!!!}

%\TODO{Brief intro of autonomous vehicle and importance of autonomous vehicle}
Autonomous vehicles shed light on the development of smart transportation system as they can improve the safety, mobility and energy efficiency, as shown in the report issued by USDOT~\cite{smith2015benefits}. One critical feature of autonomous vehicles, comparing to their human counterpart, is that they have a very short response time between an event and the vehicle reaction. The processing time for the perception-decision-response process for the human-driven vehicle is determined by human response time, which can be affected by driver's distraction~\cite{laberge2004effects}, age~\cite{caird2007effect} and the possibility of DUI ~\cite{yadav2019modelling}. For autonomous vehicles, the computer-vision-based system has a shorter response time~\cite{9632370}, and the reaction time is mainly determined by the mechatronic properties of the vehicle itself. 

%\TODO{vulnerability of autonomous vehicle in perception: current work}
Given the fact that autonomous vehicles are safety-critical cyber-physical systems, there has been a lot of research work targeting the vulnerabilities within the system. Lot of research on vulnerabilities in computer vision ~\cite{goodfellow2020generative,goodfellow2014explaining,szegedy2013intriguing} is also applicable for launching an attack on perception module of autonomous vehicles~\cite{8968267,9283977,eykholt2018robust}. %\TODO{cite adversarial attack on an autonomous vehicle here based on application: object detection, lane centering, MOT}. 
%Note that all the current works targeting the perception module of autonomous vehicle care about accuracy only. 
Generative adversarial examples can also enable generating perturbations that slowdown the inference speed of deep-learning modules, as shown in ~\cite{chen2022nicgslowdown,haque2020ilfo,hong2020panda,chen2023slothbomb}. For example, if a neural network normally inference around 0.1 second(which is a common standard for autonomous driving perception system), when given some attack data, some neural networks' inference speed can be greatly reduced. As our work shows in section \ref{sec:end2end_delay}, the vitim model will have an average inferece time around 0.1 seconds. But, when under attack its inference time goes up to 3 seconds. This means when the inference process finished, the vehicle though it is responding to the most current environment state, which was actually happened 3 seconds ago. This is way too slow for safety-critical cyber-physical system like autonomous vehicles. %This work aims to evaluate the effect of such inference time attacks on vehicle performance.
%\TODO{What did we do in this work: perform an initial analysis of the impact of inference time attack on autonomous vehicle's perception system}
In this work, we perform an initial impact analysis of inference time attack on autonomous vehicle's perception system. As the first target, we apply such an attack towards an autonomous vehicle at a signalized intersection, since in such scenario, the perception system of the autonomous vehicle not only need to detect other road users, but also the traffic signals. %~\cite{zhu2021safety}. 
%In such scenario, we have a human driving vehicle on the main highway with higher free-flow speed and the autonomous vehicle is the merging vehicle on-ramp. %\TODO{why we chose ramp merging here, BRIEF!}

% \TODO{Attacker's requirement and assumption}

% \TODO{Something for figure here!!!!!!}
% \begin{figure}
%     \centering
% \includegraphics[width=0.5\textwidth]{Figure/Scenario_figure_all_chap2.pdf}
%     \caption{The whole big picture of the assumption in this work}
%     \label{fig:assumption_big}
% \end{figure}

%\TODO{attack goal, benefit, etc.}
With such a threat model, the attack goal in this paper is to maximize the degradation of the performance of an autonomous vehicle in the intersection navigation scenario. The attacker's primary target is to downgrade the autonomous vehicle's safety performance while also seeking to compromise mobility and comfort. To achieve the attack goal, smart selection of attack policy is essential, which is the main contribution of this paper.

The contribution and novelty of our work are shown below: 

\begin{itemize}
    \item We are the first to study the effect of inference time attacks on the perception module of autonomous vehicles. Specifically, we find that the optimal attack policy requires a careful selection of both attack intensity and the launch time. This attack is different from the DoS attack occurring in V2V communication shown in previous work ~\cite{wang2020modeling}, as the information retrieval from the environment is not completely blocked out and we also involve the prediction module in state-of-the-art middleware system, which is designed to mitigate the information retrieval delay from perception.
    \item We performed end-to-end evaluation on both inference time only and performance from vehicle side and transportation engieering side. We showed that such attack can threaten both the ego vehicle and other traffic participants as well.

    % \TODO{Not sure about technical novelty}
    
    %\item \textbf{Technical Novelty.} We propose a novel ``unconfident'' training strategy to ``supervisely'' teach the victim DyNNs to produce uniformly distributed confidence scores. After injecting the backdoors to the DyNNs, the DyNNs will produce uncertain predictions for triggered inputs, forcing the DyNNs to continue computing without early termination. 
    % \TODO{Need adjustment here, no ramp merging!}
    % \item \textbf{Evaluation.} We evaluate the effect of inference time attacks  on the perception module for the autonomous vehicle. We also apply such a setting in the ramp merging scenario with various settings (details could be found in section \ref{sec:attack_analysis}). The evaluation results show that both the delayed inference time and the selection of attack launch time are important to effectively downgrade the performance of the victim vehicle. 
\end{itemize}

The structure of this paper is as follows: Section \ref{sec:LR} shows current work regarding autonomous vehicle perception time and inference time attack. Section \ref{sec:threat_model} shows the threat model in this work, mainly about attacker assumption and attacker's goal. Section \ref{sec:attk_formulation} shows how the attack is formulated and how adversarial example is obtained. Section \ref{sec:eval} shows the impact analysis with end-to-end simulation system, on inference only, vehicle and transportation side. Section \ref{sec:discussion} shows potential future work with regard to defense, and section \ref{sec:conclusion} concludes this work.

\section{Literature Review}\label{sec:LR}
% \TODO{From VehicleSec ver, will adjust later}

\begin{figure}
    \centering
    \includegraphics[width=0.8\textwidth]{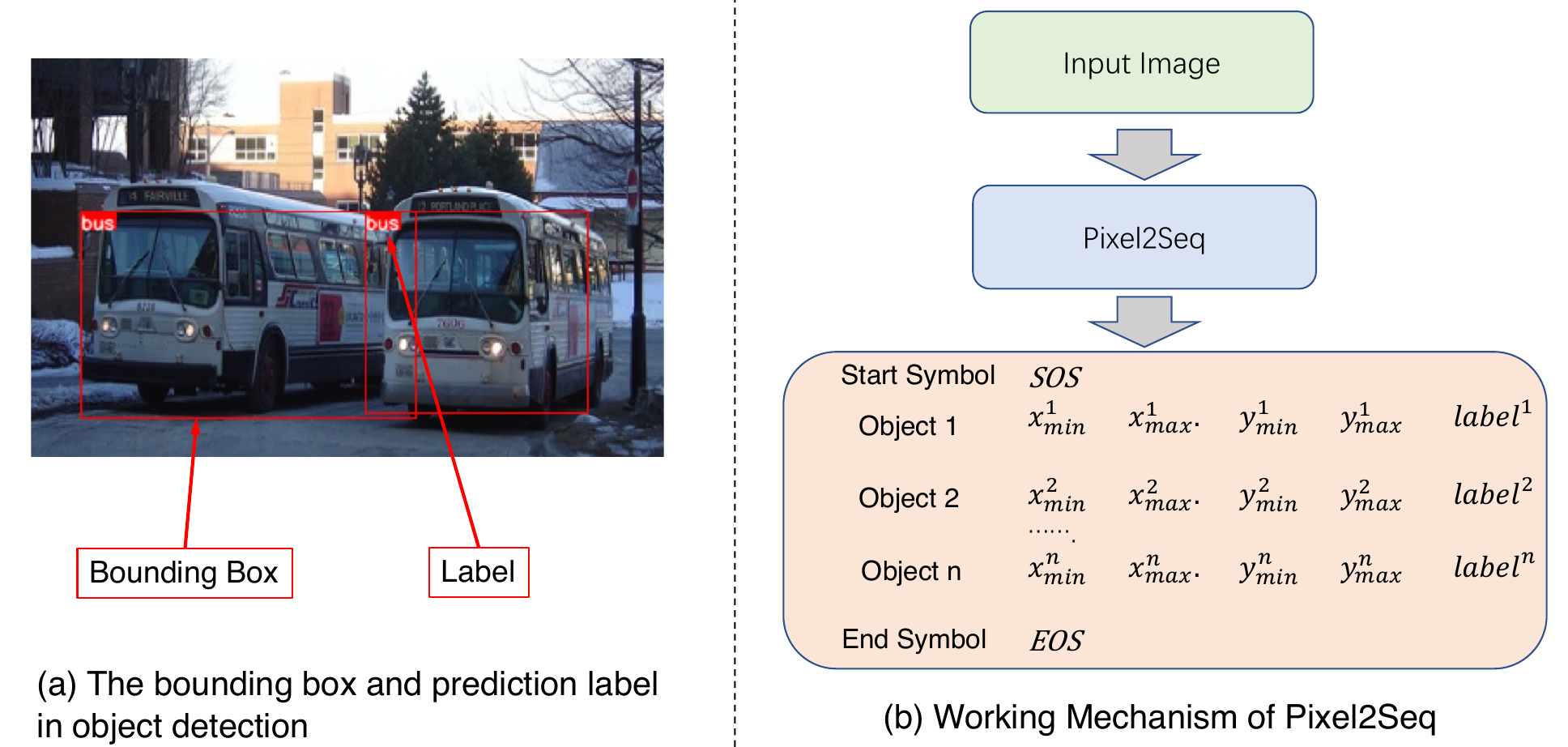}
    \caption{Problem definition of Object detection and the working mechanism of Pixel2Seq}
    \label{fig:pixel2seq}
\end{figure}

\subsection{The perception and control of autonomous vehicles}
\label{sec:object}
%\TODO{Defn of autonomous vehicle, the importance of perception}\\
Autonomous vehicles use the information obtained from sensors to make control decisions. 
% The short inference time from computer vision enabled models to outperform human-driven vehicles in terms of object, event, detection, and response \cite{9564972,cao2020highway} \CM{I do not understand the above sentence}. 
In this paper, we focus on attacking the object detection perception module for the following reasons: \textit{(1)} the object detection module plays a vital role in autonomous vehicles; \textit{(2)} the object detection module is the most widely deployed in autonomous vehicles~\cite{feng2021review}.

As shown in Figure \ref{fig:pixel2seq} (a), given one input image, the object detection module seeks to output the bounding box and the class labels of each object in the image.
Here, we briefly introduce several working mechanisms of the neural networks-based object detection module.
The first type is R-CNN, the masterpiece of two-stage object detection neural networks. In the first stage, R-CNN applies its region extraction neural networks to generate object region proposals.  In the second stage, R-CNN uses its feature extraction neural networks to extract features for each extracted object region and uses an SVM to classify the category of each region.
The second type is YOLO, which combines the stage of object region proposal and region classification into one neural network. YOLO applies only one network to divide the input image into regions and predicts bounding boxes and probabilities for each region.
Recently, state-of-the-art, Pixel2Seq, applies a language model (LM) to cast the object detection task. Object descriptions (\textit{e.g.,} coordinate of the bounding boxes and class labels) are expressed as sequences of discrete tokens, and the neural networks are trained to generate the desired sequence.
Figure \ref{fig:pixel2seq}(b) shows the working mechanism of Pixel2Seq. The backend neural networks start with the start symbol (i.e., SOS) and iteratively compute the coordinate of the bounding boxes and class labels for each object until the output reaches the end symbol (i.e., EOS).
From the working mechanism of Pixel2Seq, an important observation is that Pixel2Seq will not stop computation until its output is EOS.
Such property brings in a new vulnerability: the attacker can craft an adversarial example to make the output of Pixel2Seq not reach EOS and increase the response latency.

For autonomous vehicles, we inherit the assumption that for perception inference time, the inference interval is about 0.1s. The such setting had been applied in a previous study for information retrieval \cite{wang2020modeling} and also in the University of Michigan's Safety Pilot Model Deployment program for data acquisition\cite{bezzina2014safety}, which is way shorter than human drivers' perception response time \cite{caird2007effect}. Based in that assumption, Pixel2seq is possible for application on AV or ADAS systems as it can performance pretty well considering both inference time and inference accuracy. We show in section \ref{sec:end2end_delay} for baseline system's inference performance and section \ref{sec:CF_end2End} for baseline end2end system response. Therefore, based on the two baseline evaluations, our target model can be used when not under attack.

\subsection{Inference time attack in CPS system}
% \TODO{Simin, please write inference time attack LR here, wrt Prof Ning Zhang's slow lidar and our EfficFrog and stuff }

In recent times, a remarkable emergence of inference time attacks has drawn significant attention, as these attacks are specifically designed to rigorously assess the robustness of cyber-physical systems (CPS). The sensitivity of CPS to latency makes these inference time attacks particularly concerning, as they can lead to serious consequences with far-reaching implications.

Inference time attacks \cite{liu2023slowlidar, hong2020panda} involve the deliberate crafting of adversarial samples, aimed at consuming significantly more computational resources from the victim system. As a result, these attacks cause a substantial increase in the system's response time, introducing delays and disruptions that can have critical impacts on real-world scenarios.
The potential ramifications of such attacks are extensive, impacting various sectors where CPS plays a crucial role. For instance, a prominent example of an inference time attack is SlowLiDAR \cite{liu2023slowlidar}, which targets the LiDAR-Based Detection module in CPS. SlowLiDAR has been shown to increase the victim's LiDAR system response latency by an alarming rate of up to more than 3000\%. This escalation in latency could be detrimental, especially in applications like autonomous vehicles, where timely and accurate sensor data processing is essential for safe navigation.
Another noteworthy example is EfficFrog \cite{chen2023dark}, which focuses on attacks against the dynamic image reception module in CPS. Unlike some other attacks that compromise accuracy, EfficFrog cunningly impacts only the victim's inference latency while having a minor impact on its accuracy. This stealthiness enhances the attacker's ability to exploit vulnerabilities without raising immediate suspicion, making it a potent threat.

Given the pivotal roles that Cyber-Physical Systems (CPS) play in various domains, the potential consequences of inference time attacks can indeed be far-reaching. Consequently, evaluating the robustness of CPS under such attacks holds tremendous significance.

%\TODO{MPC in autonomous vehicle}\\
% Various control methods have been proposed to control autonomous vehicles based on perception information. Among them, \textbf{Model Predictive Control} is commonly used. Generally speaking, Model predictive control executes the planning process in a rolling horizon manner and can take the change in the surrounding environment into account in a unified framework \cite{xu2019design,xu2021system,liu2017path}. In the ramp merging scenario, MPC is used as a lower-level controller for connected and autonomous vehicles \cite{fang2022ramp}, which is the reason that we chose to implement such a controller in this work.

\section{Threat Model}\label{sec:threat_model}
% \TODO{Simin, Wenyu and Professor Yang, please write the threat model formulation here. Sticker attack only if billboard attack did not finish that's fine.}

\subsection{Attacker's Goal}
In this research, our primary objective is to investigate the manipulation of the inference time of a targeted autonomous vehicle's (AV) perception module, which can have serious implications on its security \cite{liu2023slowlidar, chen2023dark}.
Our study revolves around the creation of adversarial perturbations specifically designed to increase the inference time of the AV's perception module, potentially leading to disruptions in its essential functions.
The introduction of adversarial perturbations to the AV's perception module aims to impede its accurate and efficient processing of incoming data. This could result in delays in decision-making, compromising the overall safety and reliability of the AV's operations.
The impact of such attacks can directly lead to several safety hazards in real-world scenarios: (1) Vehicle collisions: If the inference time of the targeted AV increases suddenly, it may not be able to apply timely braking or make appropriate maneuvers, increasing the risk of collisions with other vehicles or obstacles, and (2) Violation of traffic rules and public security issues: The compromised perception module might fail to detect traffic signals, pedestrian crossings, or road signs, leading to the AV breaking traffic rules and posing security risks to the public.

\subsection{Attacker's Assumption} 

In this work, we assume that the attacker has some knowledge about the autonomous vehicle's perception module and has already given a set of trojan triggers with different attack intensities. In this specific scenario, the stronger attack intensity refers to the longer inference time delay caused by the attack. The attacker can also launch the attack with a selected starting time. For example, place the trigger on a billboard or put it as a sticker on the side of the road. We inherit the attack assumption in work ~\cite{chen2023slothbomb}. Once the attack intensity is determined, the attacker can use a universal trojan trigger. Thus the attacker does not have to generate different perturbations given each frame of perception data. 

Our  primary focus lies within the white-box attack setting, wherein the attacker possesses complete knowledge of the perception module utilized in the target Autonomous Vehicle (AV) systems. This assumption of having full access to the system's internals is consistent with prior adversarial attacks on AVs relying on camera or LiDAR technology \cite{liu2023slowlidar, cao2021invisible}.
To achieve this white-box scenario, the attacker can acquire a victim AV through purchase or rental and subsequently engage in reverse engineering of its perception module. The feasibility of such reverse engineering has been demonstrated in the case of Tesla Autopilot, indicating the plausibility of accessing and comprehending the inner workings of the AV's perception module.
By exploring the white-box attack setting in this manner, we aim to gain deeper insights into the vulnerabilities of AV systems and devise more robust defense strategies to enhance their security against potential adversarial threats.

In addition to the aforementioned scenarios, we also consider the possibility that the attacker can covertly paste an adversarial perturbation on the victim's camera. This assumption is consistent with existing research and holds significant practical relevance in real-world situations \cite{song2018physical, komkov2021advhat}. For example, the attacker could secretly affix a sticker or a specially crafted object to the camera when the victim vehicle is parked in a parking lot or any other vulnerable location.

\section{Attack Formulation}\label{sec:attk_formulation}

\subsection{Problem Formulation}

This paper is focused on attacking the perception modules of autonomous vehicles, with a specific target being the state-of-the-art object detection neural networks such as Pixel2Seq. Models such as Pixel2Seq formulate the object detection problem as a sequential generation task and employ neural networks to accomplish this task.
The key characteristic of the sequential generation model is its iterative generation of discrete tokens to represent bounding boxes and class labels for each detected object. The process continues until the end of the sequence (EOS) symbol is reached, indicating the end of the generation. In this kind of model, the inference process will not terminate until EOS occurs, thus our attack aims to minimize the likelihood of the EOS symbol, compelling the model not to terminate prematurely. In essence, we want to prolong the object detection process to disrupt the network's performance.
Formally, our attack can be formulated as the following optimization problem:

\begin{equation}
\centering
\label{eq:define}
    \begin{split}
        & \quad \Delta = \text{argmax}_{\delta} \quad \text{Latency}_{\mathcal{F}}(x + \delta) \\ 
        & s.t. \quad ||\delta|| \le \epsilon \; \wedge \; ||x + \delta|| \in [0, 1]^n \\
    \end{split}
\end{equation}
where the $x$ in the original input captured by the camera,
$\delta$ is the optimal adversarial perturbation under solving, $\text{Latency}_{\mathcal{F}}(\cdot)$ is the function that measures the object detection neural networks latency. The constraint $||\delta|| \le \epsilon$ limits the size of the adversarial perturbation and makes the perturbation unnoticeable, and $||x + \delta|| \in [0, 1]^n$ limits the adversarial examples should be a realistic one. Following existing work \cite{szegedy2013intriguing}, we set unnoticeable perturbation size constraint $\epsilon$ as 0.03; such a setting can ensure the adversarial perturbations are unnoticeable.

% \subsection{Distinction to Existing Correctness Attacks} Although several existing adversarial attacks target the perception modules of victim AVs, our work distinguishes itself by focusing on Correctness Attacks from a different perspective. Existing correctness-based attacks aim to make the model produce incorrect predictions, whereas our attack's goal is to increase the inference time of the perception modules.
% Notably, our attack presents a more challenging task compared to traditional adversarial attacks. In conventional attacks, obtaining the adversarial perturbation is relatively straightforward as the model's predictions can be directly obtained, allowing the use of gradient-based algorithms for optimization. However, in our attack scenario, the inference latency of the victim model is not differentiable, making it infeasible to use gradient algorithms to search for the optimal perturbation.
% The non-differentiable nature of the inference latency poses a significant challenge in crafting adversarial perturbations. Unlike conventional attacks, where gradients can guide the perturbation search, our approach requires alternative strategies to manipulate the inference time effectively. This unique characteristic of our attack underscores the importance of developing novel techniques and methodologies to achieve the desired impact on the perception modules' inference time.

\section{Evaluation}\label{sec:eval}
\subsection{Simulation System Setup}
% \TODO{less than 1P, describe system structure and info flow. Copied from NDSS ver, need adjustment here!!!}

Two kinds of vehicles will be discussed in all scenarios, Vehicle Under Test (VUT) and Background Vehicle (BV). VUT here refer to the autonomous vehicle, and BV refer to other traffic participants that can have impact and interact with VUT. In this paper, VUT is an autonomous vehicle where all information about surrounding vehicles is detected from its perception module.

The evaluation platform consists of a multi-resolution simulation system composed of a perception module, a background vehicle controller, and middleware. In this system, the perception module relies on the CARLA simulation, which generates perception data for the VUT. The movements of the background vehicle and VUT are reflected in CARLA, impacting the corresponding scenarios and changing the perception data. The background vehicle is controlled by SUMO \cite{8569938}, a microscopic traffic simulator, which follows traffic rules and generates traffic signals in the traffic network.

In this work, we use Robot Operating System(ROS) as a standardized middleware, providing the functionality of generating vehicle planning and control given the inferred perception information. We chose ROS because it is widely applied in autonomous driving systems such as Autoware \cite{autoware} and Apollo \cite{Apollo}. We implemented the perception, prediction and control module within ROS to test the impact of inference time attack on both vehicle side and transportation side. The Pix2seq algorithm is implemented as the inference model for perception, taking perception data from CARLA and outputting the inference information for both background vehicles and traffic lights. This inference information serves as the input for the VUT middleware system. 

For evaluation, we let VUT interact with background vehicle and traffic signal without launching inference time attack. This is to indicate that the functionality of perception, prediction and control module we implemented is fine. Then we launch the inference time attack with the same background setting for VUT, trying to see the impact of attack.
% \TODO{Need a sentence to explain the overall architecture. please also describe the input and output of each modules.} 
A figure of the evaluation platform is shown below. %and detailed description of vehicle planning and control algorithms will be shown next.
\begin{figure}[h]
    \centering
    \includegraphics[width=0.8\textwidth]{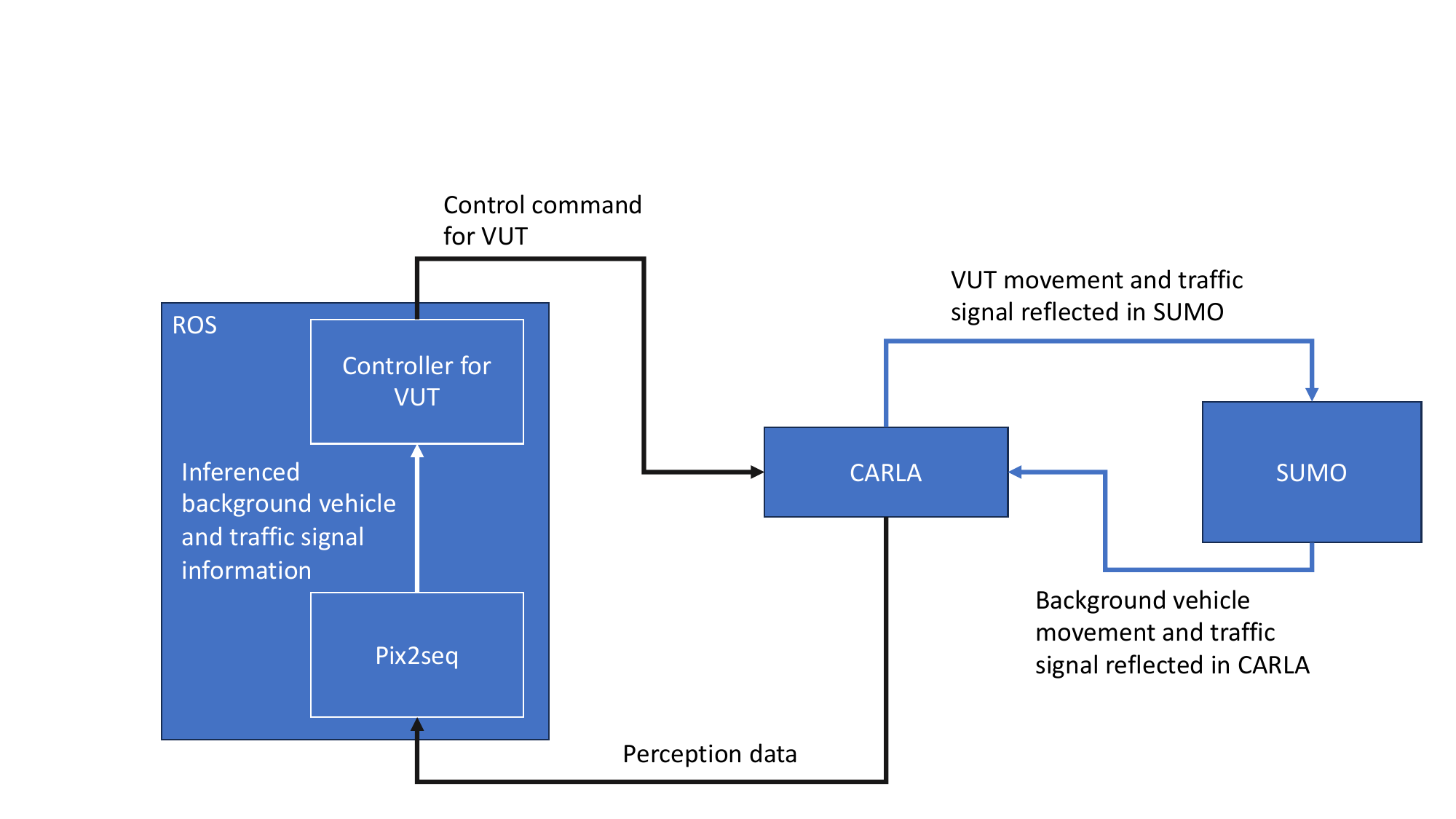}
    \caption{System architecture for end-to-end evaluation}
    \label{fig:system_end2end}
\end{figure}
\subsection{Experiment Setup}
% \TODO{describe Town04 and merchenizh rather dataset}

%%%%Evaluation platform
Following other works relate to AV cybersecurity, In our experiment, perception data is generated in CARLA and background vehicle is generated and controlled by SUMO. Two simulators are synchronized and the AV in experiment is controlled by ROS where an optimal controller is used to generate longitudinal vehicle speed control command. %For evaluation in Merzenich, all background vehicle will act cooresponding to the collected neutralistic driving data(NDD) from ExiD dataset \cite{exiDdataset},
All background vehicles will follow the IDM car-following model \cite{treiber2000congested}. Parameters of the IDM car-following model is determined by SUMO. All background vehicle will disable the emergency collision avoidance function. 

% \begin{figure}[h]
%     \subfloat[Evaluation Scenario: Town04]{\includegraphics[width=0.45\linewidth]{Figure/Town04.jpg}
%     \label}\hfill 
%     \subfloat[Evaluation Scenario: Merzenich]{\includegraphics[width=0.45\linewidth]{Figure/merzenich_rather.png}\hfill 
% \caption{Snapshot of two evaluation scenarios used for both perception evaluation and end to end evaluation}~\label{fig:mu_comparsion}
% \end{figure}

\begin{figure}[h]
    \centering
    \includegraphics[width=0.8\textwidth]{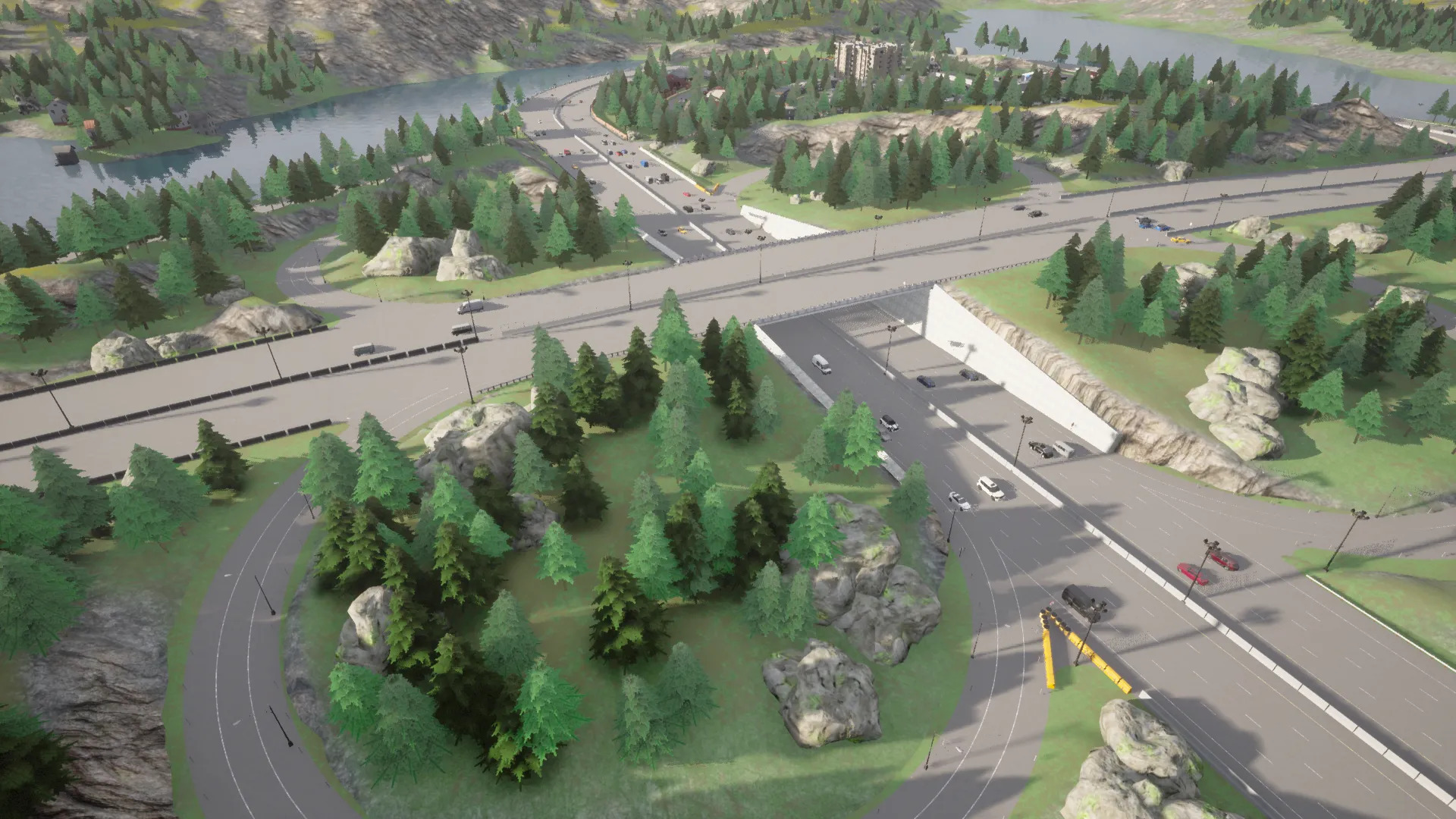}
    % \subfloat[Evaluation Scenario: Merzenich]{\includegraphics[width=0.45\linewidth]{Figure/merzenich_rather.png}
    % \label{fig:merzenich_rather}}\hfill
\caption{Snapshot of evaluation scenarios used for both perception evaluation and end to end evaluation}~    \label{fig:Town04}
\end{figure}

All scenarios are reconstructed/build in CARLA simulator and the perception data is also generated by CARLA.The evaluation scenario is in Town04 map provided by CARLA simulator as shown in figure \ref{fig:Town04}.
%, the other scenario is from ExiD dataset \cite{exiDdataset}, where we chose the testing location in merzenich rather, as shown in figure \ref{fig:merzenich_rather}. 

For vehicle control, we implement a model predictive controller (MPC) for VUT, which is a widely used control model for autonomous vehicles \cite{9564942,liu2017path}. A detailed modeling of vehicle controller will be discussed in section \ref{sec:VUT}.

\subsection{Description of background vehicle and vehicle under test}\label{sec:VUT}
 We take the assumption that the VUT observe surrounding vehicle though pixel2seq\cite{DBLP:journals/corr/abs-2109-10852}, and the relative coordinates of BV is obtained along with corresponding depth information. We assume that the depth information given to VUT is correct, as error in depth information does not contribute to the error introduced in the system. We then obtain the relative coordinate from the 2d front-person-view data and convert to relative coordinate with regard to ego vehicle. Given multiple cameras are mounted on VUT, redundancy reduction is performed to lower the impact of duplicated objects on the downstream application of prediction and planning. Based on the relative coordinates after redundancy reduction and the VUT's ego status, the relative coordinates of background vehicles obtained from perception is then obtained and used as inputs for VUT's prediction, planning and control. 

For VUT, the planning and control decision is made based on the observed BV information. If BV information has delays according to the system time stamp, then the possible BV location will be calculated by the prediction module within the controller. The prediction module had been integrated with the controller. The controller formulation of VUT is described below.

% \TODO{The following model is from VehicleSec ver, edit it!!!!!}\\
In this scenario, vehicle $n$ is the background vehicle traveling on the highway main lane and may interact with VUT thus VUT had to respond to it using perception information. Vehicle $i$ is the autonomous vehicle under investigation, also the victim vehicle in this work. The control (i.e., acceleration) of vehicle $i$ is determined by a nonlinear optimal controller described later. 

The status of the scenario is described by vehicle $n,i$. As we assume that all vehicles are traveling at the center of the lane, no lateral movement is considered. Thus we use 1-D location $x$ and longitudinal speed $v$ to describe the status of each vehicle. For each vehicle involved in the system, we denote the vehicle status as $s^k_j$, where $k$ refers to the timestamp and $ j\in \{n,m,i\} $, referring to the vehicle index. We then have $s^k_j=(x^k_j,v^k_j)$

We treat the system as a discrete-time system and use a constant speed kinematic model to describe the state transition of the system. This status update function has been used in the prediction module of Autoware.AI \cite{autoware}, which we apply in this paper. Such transition is described below as:

\begin{equation}
\begin{aligned}
    z_{k+1}=z_{k}+F(z_{k},u_{k})
\end{aligned}
\end{equation}

where $z_{k}=[s^k_n,s^k_i]$ and $F(z_{k},u_{k})=[\Delta t*v_k,0]^T$ for state estimate function from observation, note that the state estimation is based on the assumption that vehicle $n$ travels on the highway main lane. 

The longitudinal speed planning strategy for the victim vehicle is that, given perception information, it wants to arrive at the conflict point as fast as possible while maintaining a certain level of performance for safety, mobility, and comfort. We set the conflict point as the back of vehicle n with safety minimum gap as 2 meters. In this paper, victim vehicle $V_i$ generates its control command using the optimal control method, which we formulated as solving a finite-time constraint discrete non-linear model predictive control problem. The problem is solved following a rolling horizon manner. Let $N$ be the prediction horizon, and we have the following formulation for the trajectory control problem:
\begin{equation}
\begin{aligned}
    min_{u1:N}\sum_{k=1}^{N}&w_1(x_i-x_c)^2+w_2(v_i-v_i^{ff})^2\\+&w_3u_i^2+w_4((\frac{x_i-x_c}{v_i}-\frac{x_n-x_c}{v_n})^2-h)^2\\
s.t. &z_{k+1}=F(z_{k},u_{k})\\
z_k &\in \mathbf{Z}, u_k \in \mathbf{U}\\
k&=0,...,N-1
\end{aligned}
\end{equation}
% \TODO{we have two vehicles here so change the model}\\
% \TODO{explain the nlmpc controller here}
By solving this nonlinear model predictive control problem, we have the control input to the autonomous vehicle at each time stamp. $w_1(x_i-x_c)^2$ regulate the distance between the autonomous vehicle and the conflict point, and we want the autonomous vehicle to arrive at conflict point as fast as possible. $w_2(v_i-v_i^{ff})^2$ regulate the speed difference between the vehicle's speed and the free-flow speed of the road. $w_3u_i^2$ penalizes large control input to improve the comfort performance for the autonomous vehicle. $w_4((\frac{x_i-x_c}{v_i}-\frac{x_n-x_c}{v_n})^2-h)^2$ enforce the safety constraints based on the time-to-collision between the ego vehicle and the background vehicle. 
In this optimization problem, $Z$ is the feasible set of all states and $U$ is the feasible set of all control commands of the victim vehicle, which considers the acceleration and de-acceleration limits of the actuator. 

We assume that the victim's state estimation of the surrounding vehicles is based only on perception, which is common in autonomous vehicle settings.  

\begin{figure}[h]
    \centering
    \includegraphics[width=0.9\textwidth]{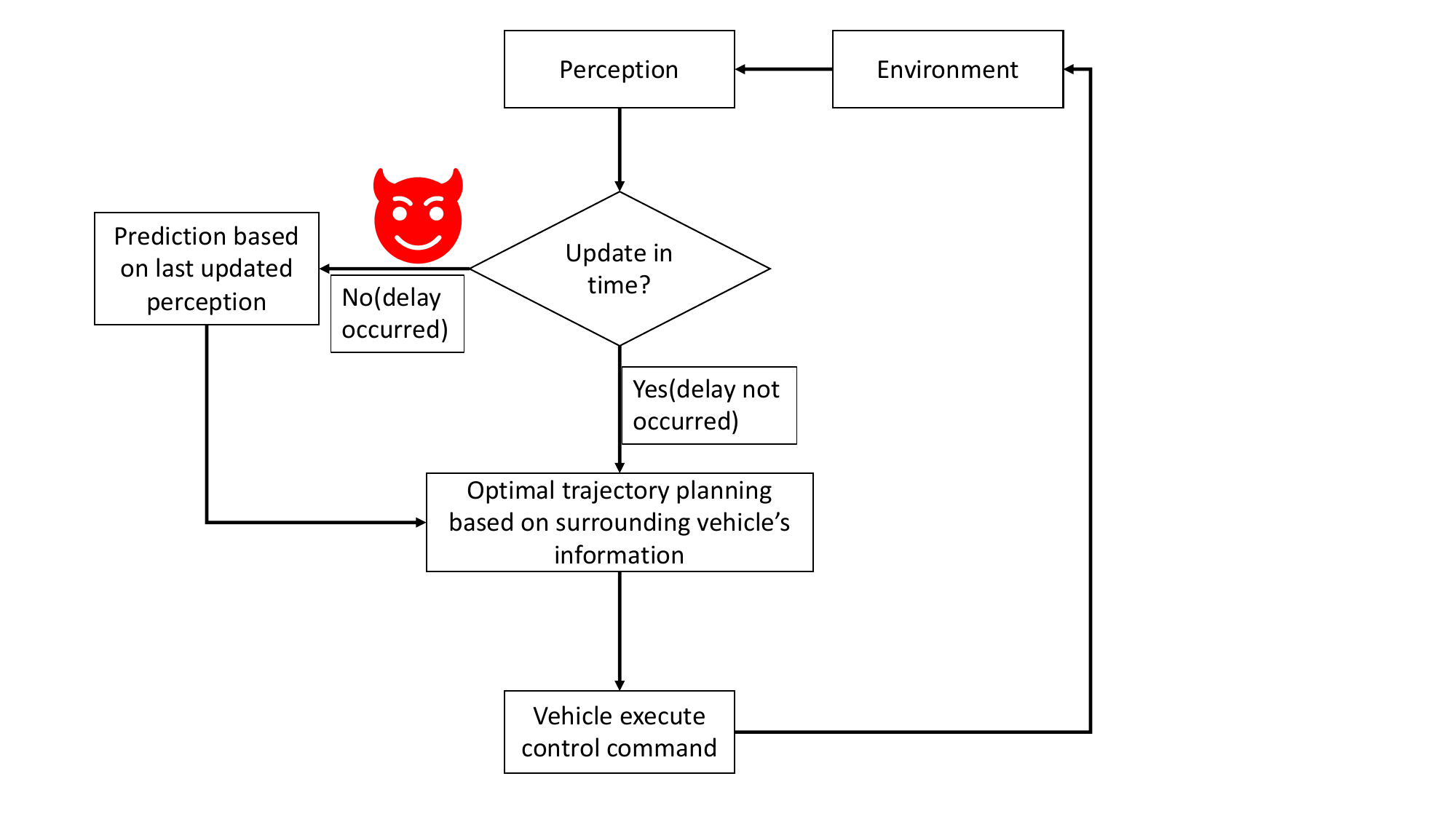}
    \caption{Information update and command generation for the victim autonomous vehicle when under or not under attack}
    \label{fig:flow_chart1}
\end{figure}

For response to traffic signal regarding autonomous vehicle, we deploy the evaluation in signalized intersection in Town04, a virtual map provided by CARLA simulator. In this scenario, we deploy a traffic signal controller at the T intersection on highway, and we apply a fixed signal time. For the highway direction, the signal time has 20 seconds green time, and 15 seconds red. 
% For VUT response, the yellow time is treated as red and they have the same response pattern. 
% \input{Eval/Attack_scenario_merge}
% \input{Eval/Attack_scenario_intersection}
\section{Result}

\subsection{Delay on End-to-end system}\label{sec:end2end_delay}

%%%%%%%%%%%%%%Already proofread by ChatGPT
Given that the impact of an inference time attack on the AV system affects both inference correctness and latency, we first evaluated the difference in inference time from an end-to-end system perspective. The reason why we perform the evaluation on inference time only at the en-to-end system at this time is that, as a safety-critical cyber-physical system, both response time and inference result has important impact on the decision making for AV system. What's more, the perception to surrounding environment with lower frequency can also result in a delayed interaction for AV with regard to surrounding environment. Even with the help of prediction,

Assuming the same initial speed and driving length, a significant difference in inference time can result in a substantial variation in the number of inferences executed. For the sticker attack, the mean inference time for pixel2seq without an attack is about 0.05-0.15 seconds, while the mean inference time for pixel2seq under attack is around 3.2 seconds. If the total driving time per run is 40 seconds, pixel2seq can perform 400 inferences without an attack, and 12.5 inferences when under attack. This significant imbalance in the number of observations makes it challenging to conduct a fair comparison.

\begin{figure}
    \centering
    \includegraphics[width=0.8\linewidth]{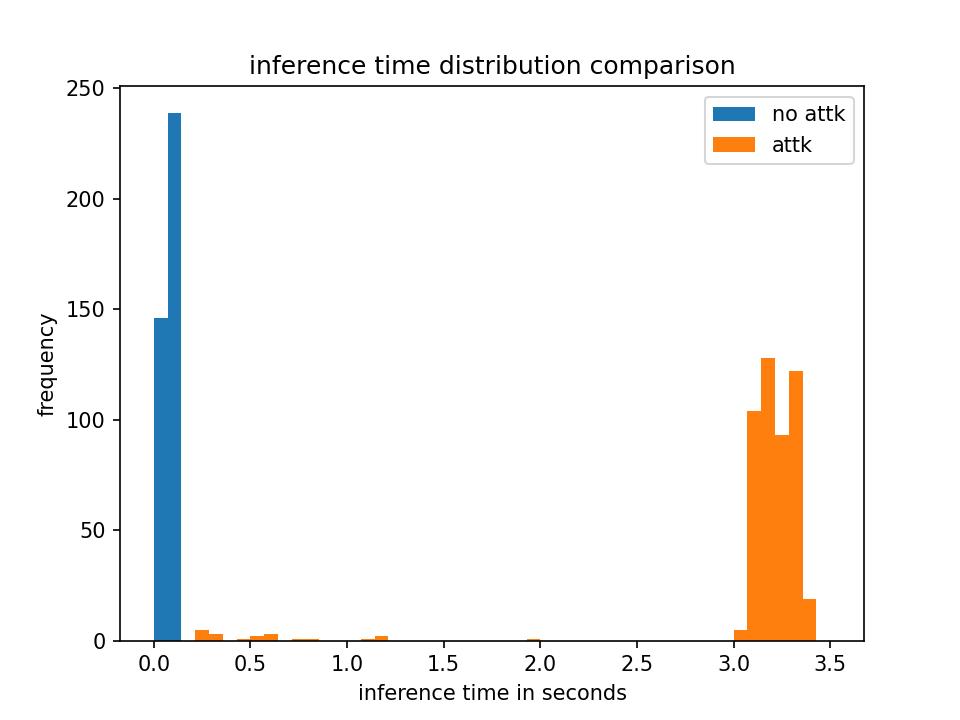}
    \caption{Histogram between baseline and pixel2seq under attack}
    \label{fig:hist_sticker_cmp}
\end{figure}

Therefore, we plotted the accumulated inference time data to demonstrate the effectiveness of the attack. The plot is shown in Figure \ref{fig:hist_sticker_cmp}. By accumulating multiple experiments for inference time attack with 40 times, we obtained the histogram for perception module when under attack. As shown in the figure, the inference time under attack is around 3.0-3.5 seconds, which is not suitable for a time-critical cyber-physical system like AV.
\subsection{Car-following case}\label{sec:CF_end2End}
% \TODO{Short version for NDSS. Elaborate here!!!!}
%%%%%%%%%%%%%%%%%%Proof read by ChatGPT
We conducted evaluations on two different scenarios and two different attack models as mentioned in section \ref{sec:threat_model}. For the car-following case with a sticker attack, we ran 40 runs and achieved a collision rate of 100\%. Moreover, upon examining the experiment log, we found that the inference process was delayed for 100\% of the frames due to adversarial perturbation. We define 'delayed inference' by comparing the inference time during end-to-end evaluation.

We observed that the inference time attack resulted not only in a delayed response but also in a falsified inference result. Both the delayed response and the falsified result violate the time-criticality requirement for AV, which is a time-sensitive cyber-physical system. The time length of each run ranged from 30-40 seconds. 

\begin{figure}[h]
    \centering
    \includegraphics[width=0.7\textwidth]{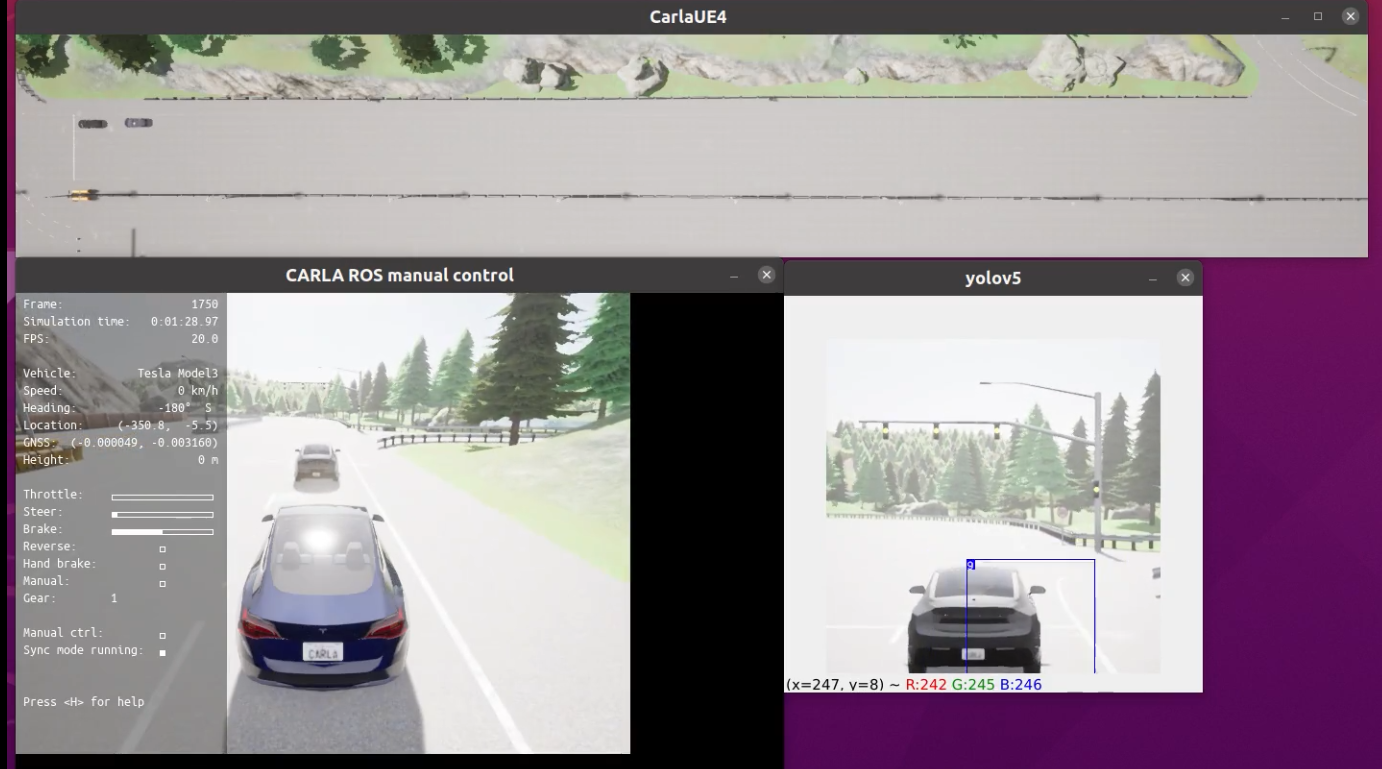}
    \caption{System snapshot and CAV perception in benign case}
    \label{fig:system_sim_good}
\end{figure}

\begin{figure}[h]
    \centering
    \includegraphics[width=0.7\textwidth]{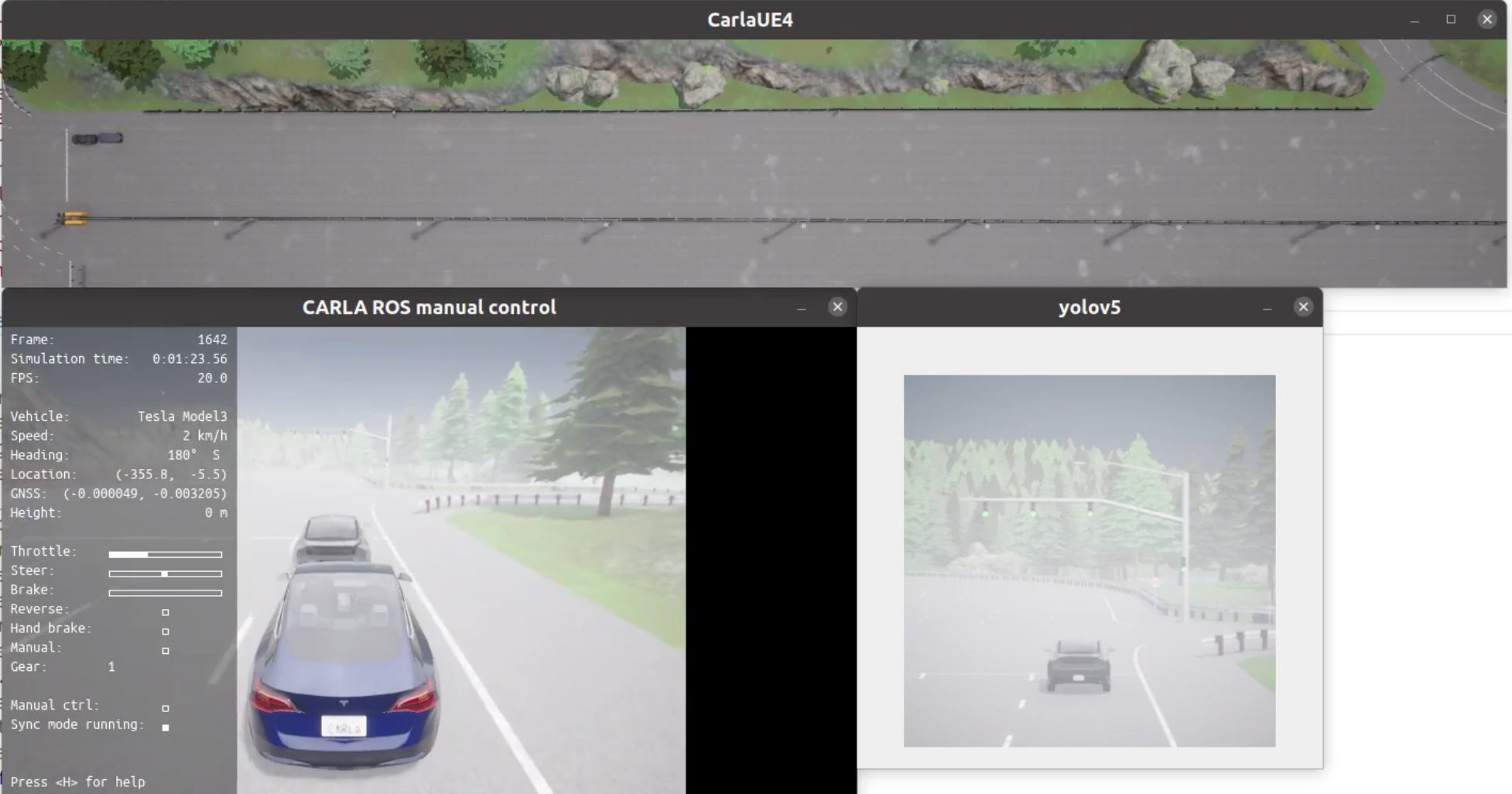}
    \caption{System snapshot and CAV perception under attack}
    \label{fig:system_sim_bad}
\end{figure}

In contrast, in benign cases, our implemented longitudinal controller always responds promptly and appropriately to the detected lead vehicle, generating a control command that ensures both safety and rider comfort. No collisions were observed in these benign cases.

We present a snapshot of our simulation system, illustrating how the Vehicle Under Test (VUT) responds to a background vehicle (BV) based on the corresponding perception result. Figure \ref{fig:system_sim_good} showcases the Bird's Eye View (BEV) of the evaluation scenario in the upper part, displaying the relative distance between the VUT and BV. In the lower left, we provide a real-time camera feed from a Front-Person-View (FPV) for the VUT, also indicating the relationship between the VUT and BV. The lower right section presents the inference and perception result from the Pix2seq algorithm. In the benign case, the BEV, FPV, and inference results from Pix2seq are consistent with each other, suggesting that no delay occurred during the evaluation scenario.

Figure \ref{fig:system_sim_bad} has a similar layout for cross-reference and represents an example of Pix2seq under attack. The BEV view reveals that a collision has already occurred, a fact also visible from the FPV that bypasses the inference. However, the lower right section, showing the inference result, still indicates that the BV is far from the VUT. This discrepancy demonstrates a "delayed inference and response," suggesting that the VUT is responding to environmental data from some time ago, thus violating the time-criticality principle intrinsic to safety-critical systems like autonomous vehicles.

% \begin{figure}[ht]
%     \subfloat[System snapshot with inference result in benign case]{\includegraphics[width=0.8\linewidth]{sim_Good.png}
%     \label{fig:CF_ts_no_collision_BV_stop}}\hfill 
%     \subfloat[System snapshot with inference result in malicious case]{\includegraphics[width=0.8\linewidth]{Sim_Bad.png}
%     \label{fig:CF_spd_no_collision_BV_stop}}\hfill
% \caption{System snapshot in both bengin and malicious case to show the impact on inference}~\label{fig:eval_sceanrio_CF}
% \end{figure}

\begin{figure}[h]
    \subfloat[Time-space diagram of both target vehicle and background vehicle in benign case]{\includegraphics[width=0.44\linewidth]{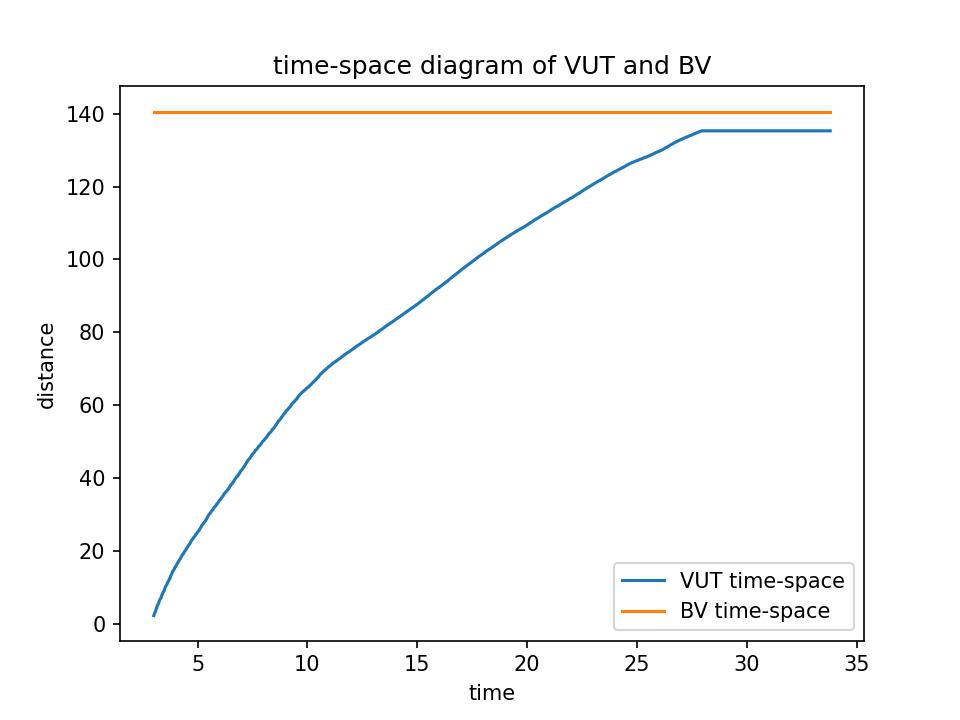}
    \label{fig:CF_ts_no_collision_BV_stop}}\hfill 
    \subfloat[Speed profile of target vehicle in benign case]{\includegraphics[width=0.44\linewidth]{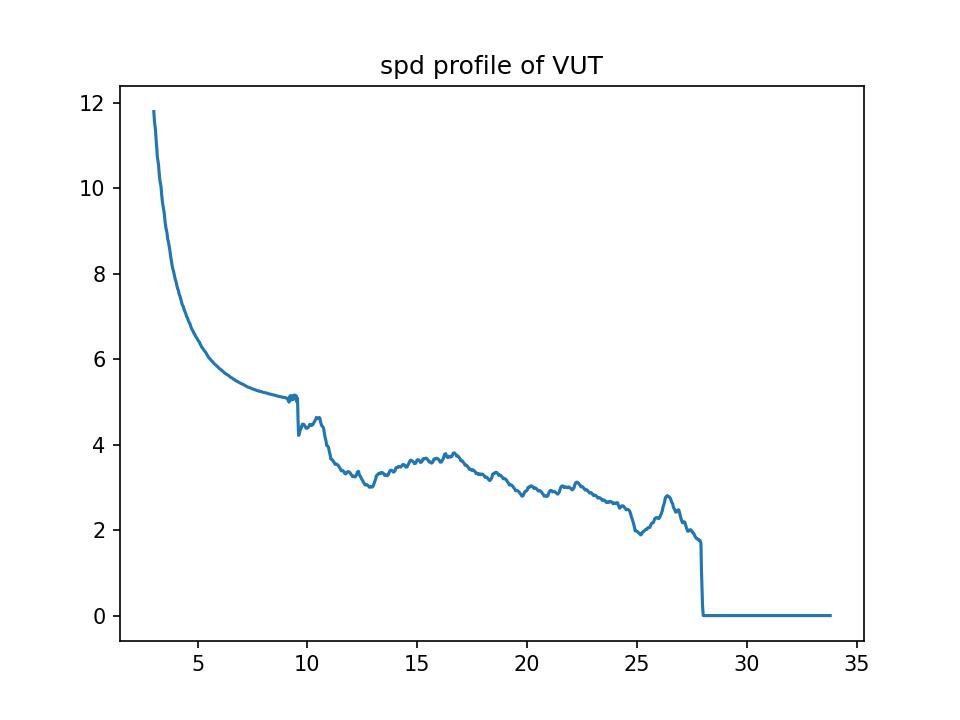}
    \label{fig:CF_spd_no_collision_BV_stop}}\hfill
    \subfloat[Time-space diagram of both target vehicle and background vehicle when target vehicle is under attack]{\includegraphics[width=0.44\linewidth]{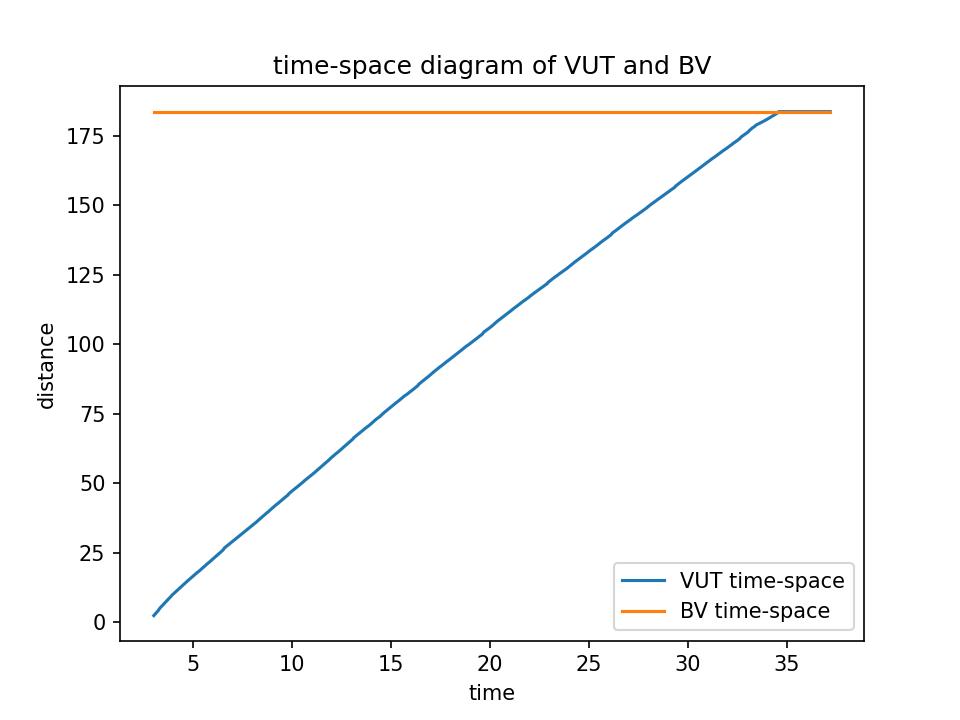}
    \label{fig:CF_ts_collision_BV_stop}}\hfill 
    \subfloat[Speed profile of target vehicle when under attack]{\includegraphics[width=0.44\linewidth]{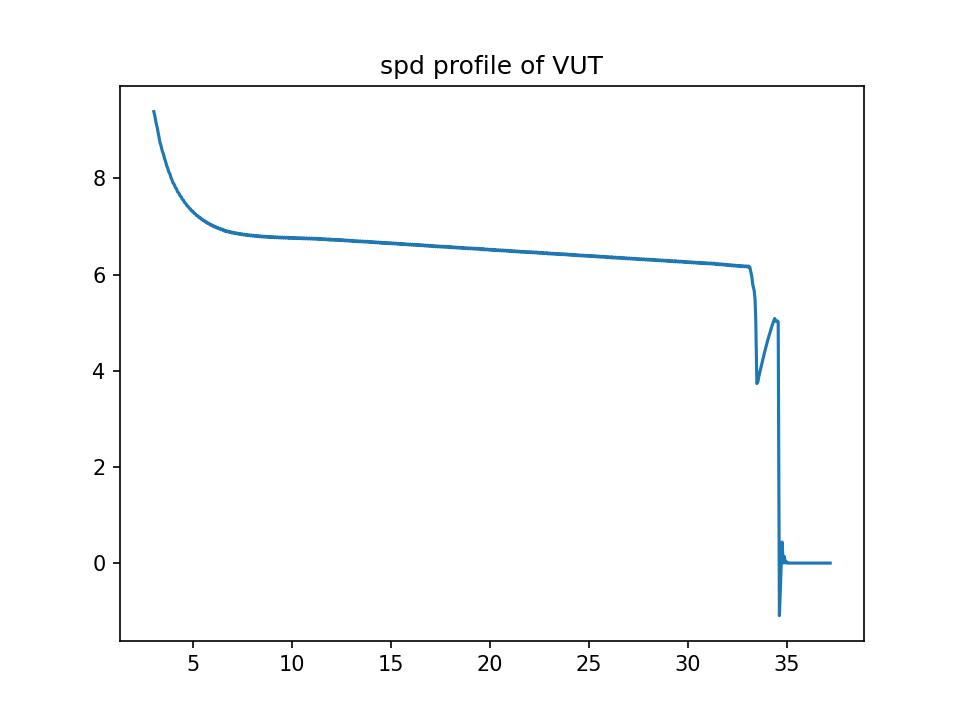}
    \label{fig:CF_spd_collision_BV_stop}}\hfill
\caption{Time-space diagram and speed profile of vehicles in benign case and under attack, when responding to background vehicle}~\label{fig:eval_sceanrio_CF}
\end{figure}

% \Purdue{It is possible that not all frames are delayed. Location of billboard can affect end2end evaluation result}
% For the pure car-following case with an adversarial billboard attack, we conducted \TODO{XXX} experiments and achieved a collision rate of \TODO{XXX}. Upon inspecting the experiment log, we discovered a \TODO{XXXXXXXXXXXX} relationship between the distance of the VUT to the adversarial billboard and the inference delay. The figure illustrating this relationship is shown in figure \TODO{XXXXX}.

%%%%%%%%%%%%%%%07132023
% \TODO{1. Typical case, BV stop, VUT respond}
For impact analysis on vehicle and the transportation system, we first evaluate a typical case where a background vehicle stay stationary in the road network, and see what is the impact of inference time attack on AV system. The time-space diagram and speed profile of victim vehicle in benign case is shown below:

% Considering the vehicle physical properties and kinematics properties, the controller will apply a brake if the planned speed is lower than a threshold. This is to avoid the impact of road geometry for the testing scenario. 

In Town04, the ground of testing scenario is not even, therefore we let the controller apply a brake instead of setting throttle and brake as zero when planning speed is zero. This caused the speed drop of target vehicle at figure \ref{fig:CF_spd_no_collision_BV_stop} around time 30s. Also the controller takes perception error from vehicle perception data , thus the estimation of the distance between the ego vehicle and front vehicle varies as the perception result itself has some error. The fluctuation of distance estimation from perception result in the planned speed for target vehicle, therefore the speed profile of ego vehicle did not showed a perfect constant speed.

We also launched the inference time attack for the same case where the background vehicle remained in the same position and tried to evaluate the target vehicle's response to the background vehicle. We obtained the time-space diagram of the VUT and the background vehicle. We also obtained the speed profile of the VUT to see the impact of the attack on VUT. Both beign and malicious cases are shown in figure \ref{fig:eval_sceanrio_CF}.

%%%%%%%%%Explain the negative speed in the speed profile

%%%%%%discussion with case that attack but no car crash

\subsection{Traffic Signal Response scenarios}
% \TODO{Elaborate for the TRB version. Get benign case for VUT responding to traffic light, red and green (red is better), get attack case for VUT responding to traffic light, red light running}

For Traffic Signal Response scenarios, the cases differ slightly from those in the car-following scenario. In traffic signal response cases, delayed perception can result in red-light-running, or delayed start-up.

\begin{figure}[h]
    \subfloat[Time-space diagram of both target vehicle and background vehicle in benign case]{\includegraphics[width=0.4\linewidth]{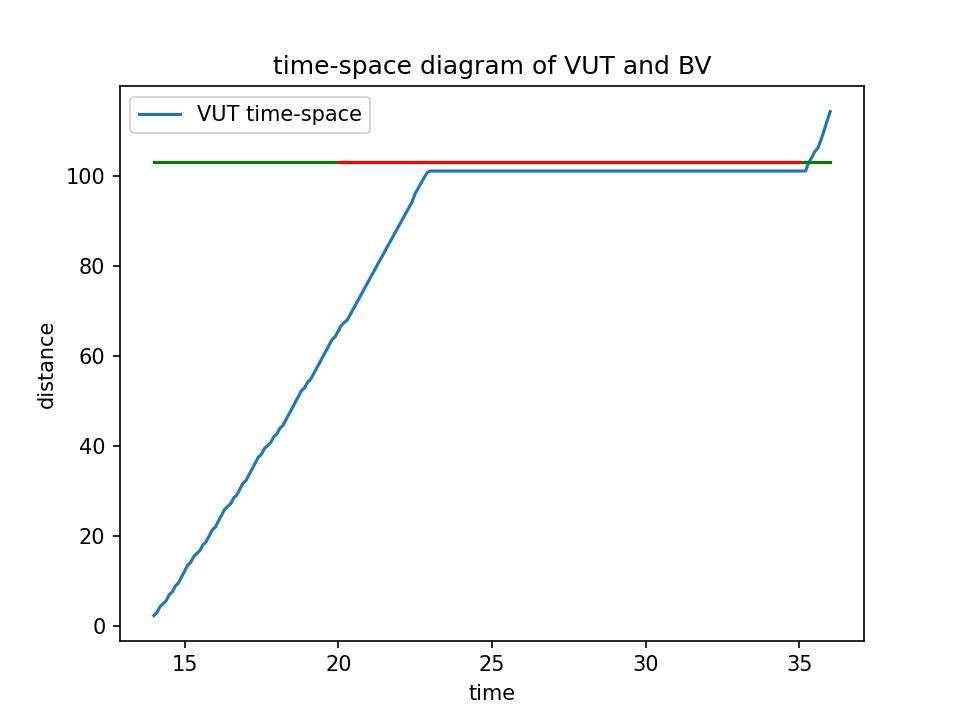}
    \label{fig:CF_ts_collision_BV_stop}}\hfill 
    \subfloat[Speed profile of target vehicle in benign case]{\includegraphics[width=0.4\linewidth]{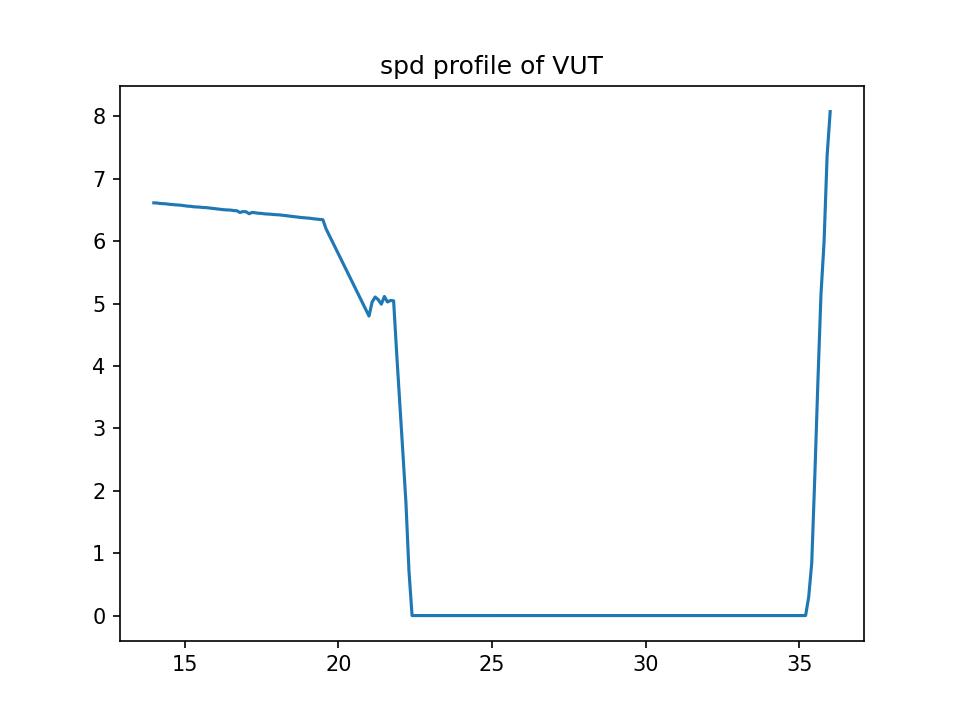}
    \label{fig:CF_spd_collision_BV_stop}}\hfill
    \subfloat[Time-space diagram of both target vehicle and background vehicle when target vehicle is under attack]{\includegraphics[width=0.4\linewidth]{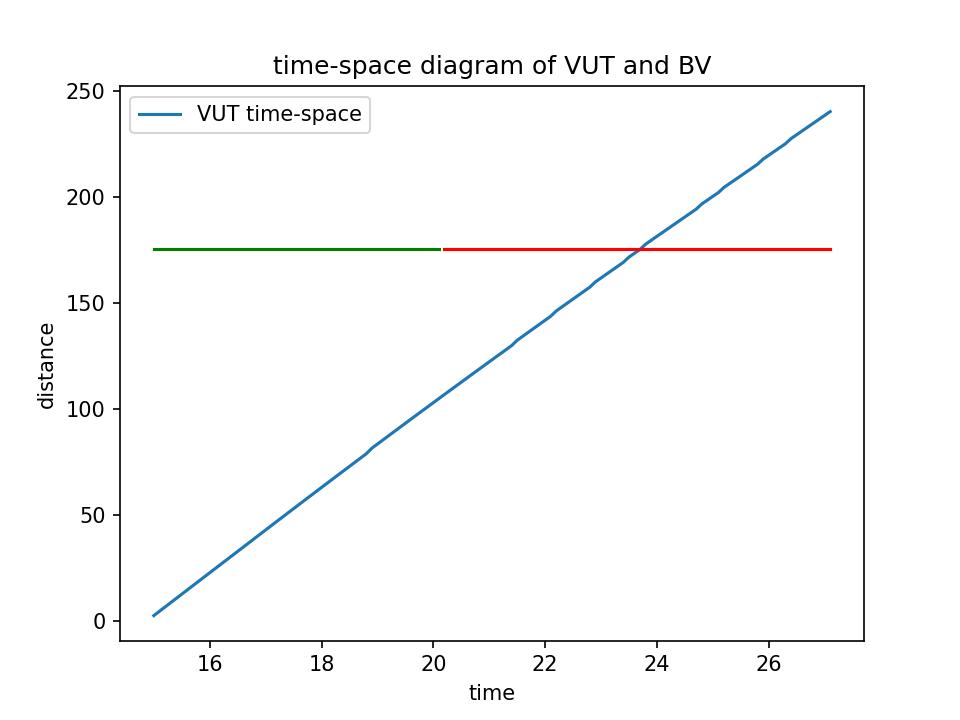}
    \label{fig:TSC_violate_TS}}\hfill 
    \subfloat[Speed profile of target vehicle when under attack]{\includegraphics[width=0.4\linewidth]{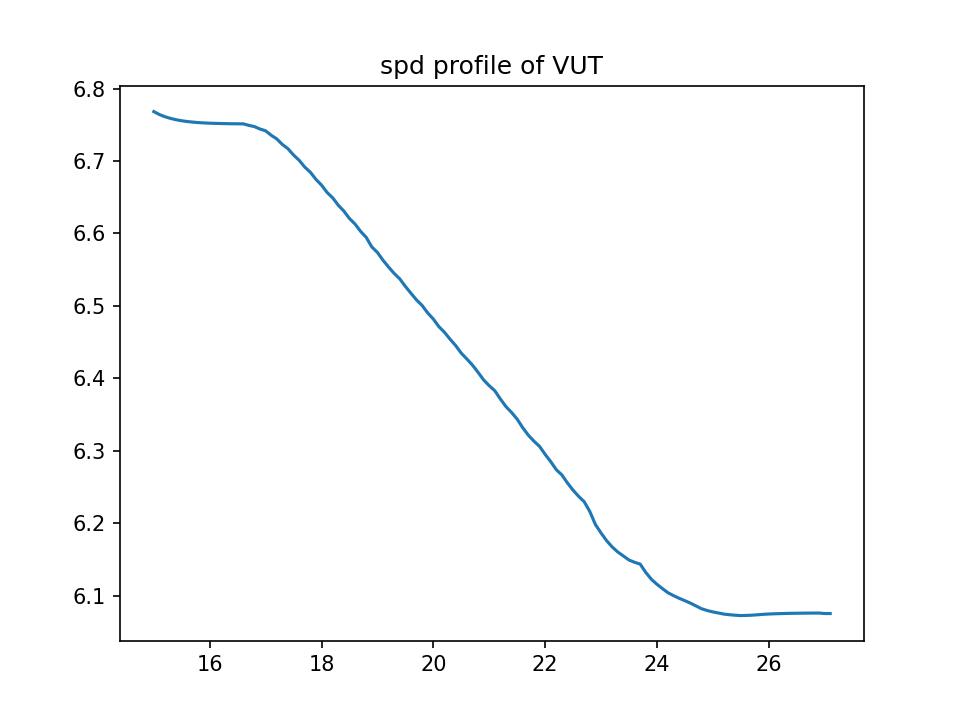}
    \label{fig:TSC_violate_Spd}}\hfill
\caption{Time-space diagram and speed profile of vehicles, benign case and under attack}~\label{fig:eval_TSC_redlight}
\end{figure}
Red-light running, which violates traffic rules and threatens safety, occurs when the actual traffic signal is red but the perception result for the VUT indicates it's green due to delayed perception. Delayed start-up mainly impact the mobility within the transportation network. This situation is the opposite of red-light running, resulting from a delayed perception when the actual traffic signal is green, but the VUT perceives it as red. It's also possible that delayed perception has no impact on the VUT's response to traffic lights. In such cases, the traffic light remains the same when the VUT passes the intersection.

For the sticker attack, we obtain the time-space diagram and speed profile of VUT in both benign and malicious cases. In the benign case, our implemented latitudinal controller could always respond properly and timely to the traffic signal once it detected the traffic lights. It generated a longitudinal control command that satisfied safety, rider comfort, and mobility requirements, considering the impact on the entire traffic network. We observed no collisions in the benign case and no violation of traffic signal. 

For malicious case, we also evaluate the scenario with time-space diagram and speed profile for VUT. As show in figure \ref{fig:TSC_violate_TS} and \ref{fig:TSC_violate_Spd}, the delayed response of VUT leads to the violation of traffic rule, as no deacceleration is performed in response to the traffic signal in red. 

% For the adversarial billboard attack, we conducted \TODO{XXX} experiments and achieved a collision rate of \TODO{XXX}. Upon inspecting the experiment log, we discovered a \TODO{XXXXXXXXXXXX} relationship between the distance of the VUT to the adversarial billboard and the inference delay. The figure illustrating this relationship is shown in figure \TODO{XXXXX}.

%%%Not prof read by ChatGPT yet

To evaluate the impact of inference time attack on AV's response to the traffic signal, we first obtained a baseline performance of AV respond to TSC system. From the baseline response of AV, we can see that the AV can respond to traffic signal correctly and stop at stop bar when it detects a red signal, and resume driving once the signal turns green. In contrast, when under attack, due to the delayed inference the VUT did not respond to the traffic signal and directly do a red-light running. Both cases are shown in figure \ref{fig:eval_TSC_redlight}.

\section{Discussion}\label{sec:discussion}

In this work, our primary focus is on the inference attack. We accept the attacker's assumption that once the attack is launched and the VUT is under attack, the inference delay is maximized. Also, in this work, the main product is the inference time delay, while the falsified inference result is a side product.

Therefore, from a vehicle and transportation perspective, we tend to consider a weaker assumption, that when an attack is launched, the delay may not reach its maximum. Additionally, we are interested in understanding if the delay in inference can have an accumulative effect on both vehicle control and the traffic state.

In this context, one possible direction for future work would be to treat the generation of adversarial samples as a sequence instead of a single adversarial sample. In other words, we'd like to explore if smartly choosing the inference delay time sequence can result in a greater impact not only on a single vehicle but also on traffic flow. The generation of such a sequence and connecting it with the generation of adversarial perturbation is one possible avenue for future work.

At the vehicle control and transportation system level, one possible defense method is to utilize the last observed perception information that has not been attacked. The current prediction method for vehicle-side control suffers from a relatively short prediction horizon. It's important to note that the common setting for the prediction horizon of the vehicle-side controller is approximately 2 seconds. However, during the inference time attack mentioned in section \ref{sec:threat_model}, based on the end-to-end evaluation, the inference delay could be as long as 3.2 seconds, which exceeds the common planning horizon.

Therefore, utilizing traffic-level prediction that has a lower update interval but longer information validity could be a possible direction. Integrating traffic-level information into the vehicle-side prediction can better equip the VUT to cope with the surrounding traffic state, and thus mitigate the attack to some extent.

\section{Conclusion}\label{sec:conclusion}

In this work, we revealed a potential threat to the AV system through perception module. We showed that certain kind of attack can delay inference of perception module in AV system and thus threaten the safety for AV. We evaluated the attack on an end-to-end simulation system using standardized middleware structure with real-time system. We performed the attack under two scenarios: response to background vehicle and response to traffic signal. In both cases we show significant difference in performance with and without attack. Future work regarding this vulnerability in inference time can relate to traffic-informed defense, and identification of AV that is under such attack. 

\section{Author Contributions Statement}
The authors confirm their contributions to the paper as follows: study conception and design: Chen, Chen,Yang, Feng; system construction: Chen, Chen, Li; data collection: Chen, Chen,Li; analysis and interpretation of results: Chen, Feng; manuscript preparation: all authors. All authors reviewed the results and approved the final version of the manuscript. ChatGPT4 is used to correct the grammar issues within the manualscript.

\newpage

\bibliographystyle{trb}
\bibliography{trb_template}
\end{document}